\documentclass{iopart}

\usepackage{graphicx}

\begin{document}
\title{The new physics of non-equilibrium condensates:
  insights from classical dynamics} 
\author{P R Eastham}
\address{Theory of Condensed Matter, Cavendish Laboratory, Cambridge CB3 0HE, United Kingdom}

\date{\today}
\begin{abstract}
  We discuss the dynamics of classical Dicke-type models, aiming to
  clarify the mechanisms by which coherent states could develop in
  potentially non-equilibrium systems such as semiconductor
  microcavities. We present simulations of an undamped model which
  show spontaneous coherent states with persistent oscillations in the
  magnitude of the order parameter. These states are generalisations
  of superradiant ringing to the case of inhomogeneous broadening.
  They correspond to the persistent gap oscillations proposed in
  fermionic atomic condensates, and arise from a variety of initial
  conditions. We show that introducing randomness into the couplings
  can suppress the oscillations, leading to a limiting dynamics with a
  time-independent order parameter. This demonstrates that
  non-equilibrium generalisations of polariton condensates can be
  created even without dissipation. We explain the dynamical origins
  of the coherence in terms of instabilities of the normal state, and
  consider how it can additionally develop through scattering and
  dissipation.
\end{abstract}

\pacs{71. 35. Lk, 71. 36. +c, 03. 75. Kk}

\section{Introduction}

Cavity
polaritons\ \cite{hopfield58,weisbuch92,skolnick98,savonapol98,kavokinbook}
are the quanta of the electromagnetic field in a semiconductor
microcavity. Since they are part photon, cavity polaritons are bosons.
Thus there is the possibility of forming a Bose condensate of
polaritons, in which an incoherent population of polaritons becomes
coherent due to the combined effects of bosonic statistics and
interactions. The last few years have seen a steady accumulation of
evidence for such physics.  Initial observations of a
threshold\ \cite{dang98,senellart99} in the intensity of the emitted
light as a function of pumping power have been supplemented by
demonstrations of both temporal\ \cite{deng03} and
spatial\ \cite{richard05} coherence in the emission.  Evidence that the
phenomenon is not conventional lasing includes the persistence of
polaritonic features in the spectrum above threshold: a polaritonic
spectrum arises from a coherent polarisation in the gain medium, while
such polarisation is negligible in conventional laser theory.

One route to developing theories of polariton condensation is to begin
with the quasi-equilibrium limit of a population of polaritons. The
established techniques of many-particle physics can then be applied to
model microcavities, and theories developed which allow for issues
such as the internal structure of the polaritons, disorder acting on
the excitons, and the many-body nature of Bose condensation.  This
approach has now established phase diagrams and observable properties
for polariton condensation in a range of increasingly realistic
models\ \cite{eastham00,eastham01,easthamthesis,szymanska02,szymanska03,marchetti04,keeling04,keeling05,marchetti05,marchetti052,eastham06}

The problem with the equilibrium theories is linking them to the
experiments, which may not be in thermal equilibrium. In principle
they are directly applicable if the polariton lifetime is long
compared with the thermalization time. Unfortunately, while the
polariton lifetime is easily established to be of the order of
picoseconds, it is difficult to get a handle on the thermalization
time. In practice the typical situation appears to be that the system
does not thermalize below the nonlinear threshold, but can do above
it\ \cite{richard05b}. Thus while equilibrium theories are apparently
sometimes directly applicable to the condensed state, further work is
needed to understand the threshold itself, and the apparent
equilibration below it. 

Polariton condensation experiments are one topical motivation for
reconsidering a fundamental issue in the physics of coherent
many-particle states: the origins of the coherence. Since in
equilibrium the coherent state is selected because it has a lower
energy than the incoherent state, one might expect the origins of
coherence to be in dissipation and thermalization. Yet coherence can
develop in the absence of dissipation (e.g. superradiance), or in
dissipative systems that are not in thermal equilibrium (e.g. lasing
and the non-equilibrium condensation proposed in\ \cite{szymanska06}).

There are many areas beyond polariton condensation where the origins
of coherence are relevant. In microcavities one has access to a range
of non-equilibrium states, created for example by coherent pumping\
\cite{baumberg00}, and hence a laboratory for widely exploring
mechanisms which create and preserve coherence. In atomic Bose gases
the development of condensates has been studied both theoretically and
experimentally\ \cite{stoof95,kohl01,davis02}, and there has been much
recent interest in non-adiabatic phenomena involving coherent states
of atomic Fermi gases\
\cite{barankov06b,yuzbashyan06b,szymanska05}.

In this paper we explore the origins of coherence by discussing the
dynamics of classical Dicke-type models. We first consider an undamped
model, which we present in section\ \ref{sec:model} along with a brief
review of some analytical results on its dynamics. In section\
\ref{sec:numerics} we then present and discuss some numerical
simulations showing that some of the coherent steady-states are
reached from some initial conditions, even in the absence of
dissipation. In section\ \ref{sec:stability} we explain, in terms of
the dynamical stability of the incoherent states, why this occurs. In
a special case of the model the results can be further understood
using exact solutions, as we discuss in section\
\ref{sec:exactsols}. In section\ \ref{sec:damping} we discuss a
phenomenological approach to adding damping to the dynamics, and
present numerical results showing the approach to an equilibrium
condensate. Finally in section\ \ref{sec:conclusions} we propose
several scenarios for how coherence could develop in a polariton
condensation experiment, and summarize our conclusions.

\section{Background and Basic Model}
\label{sec:model}

The basic model we consider here directly describes excitons strongly
localised on disorder in a three-dimensional cavity. It has the
Hamiltonian
\begin{equation}
\label{eq:spinham} H = \omega_c \psi^\dagger \psi + \sum_i \left[E_i S^{z}_{i} +
\frac{g_i}{\sqrt{N}} \left( S^{+}_{i} \psi + \psi^\dagger S^-_i \right)
\right]. \end{equation} $\psi^\dagger$ is the creation operator for a cavity
photon, with energy $\omega_c$. The dielectric is modelled as a set of $N$
two-level systems, with the $i^{th}$ two-level system described by the
spin-half operators $\vec{S_i}$. The eigenstates of $S_i^z$ correspond to the
presence or absence of an excitation on site $i$.

The form (\ref{eq:spinham}) is a generalisation of the well-known
Dicke model to include a distribution of exciton energies (i.e.
inhomogeneous broadening) and coupling strengths. It is one of the
central models of quantum optics, describing for example
lasing\ \cite{haken75}, equilibrium superradiance\ \cite{hepp73}, and
dynamical superradiance\ \cite{dicke54}. It can be viewed as a
generalization of the BCS Hamiltonian to the strong-coupling regime,
as is apparent on rewriting the spin operators in terms of two species
of fermions. It was applied to polariton condensation by Eastham and
Littlewood\ \cite{eastham01,eastham00,easthamthesis}, who considered its
quasi-equilibrium thermodynamics at fixed excitation number
\begin{equation} \label{eq:exnum} L=\psi^\dagger \psi+ \sum_i
S_i^z; \end{equation} note this is conserved by
(\ref{eq:spinham}).  The resulting polariton condensate can be viewed
as a generalization of the BCS state to include coherent
photons. Later work on the equilibrium polariton condensate in models
of the basic form (\ref{eq:spinham}) includes generalisations to
include propagating photons\ \cite{keeling04,keeling05},
decoherence\ \cite{szymanska03}, and more realistic approaches to
disorder\ \cite{marchetti04,marchetti052}. The same theoretical
framework has also been applied to condensation in atomic gases of
fermions\ \cite{holland01,timmermans01}.

To study the dynamics of the model (\ref{eq:spinham}) we begin from
the Heisenberg equations of motion for the operators. However, we
shall not consider the full quantum dynamics described by these
equations. Instead we consider the simpler problem obtained by
replacing the operators in the equations of motion with classical
variables. This is equivalent to taking the expectations of the
equations of motion and approximating the expectation values of
products as products of expectation values. Thus we consider a
mean-field dynamics for the quantum model. Since mean-field theory is
exact for the thermodynamics in the limit $N\to \infty$ we expect it
to account for many features of the dynamics when $N$ is large.

Thus the classical equations of motion we consider are \numparts
\begin{eqnarray} i\dot{\psi}&=&\omega_c \psi + \frac{1}{2\sqrt{
N}}\sum_i g_i \sigma_i^-,
\label{eq:ph}\\\dot{\vec\sigma_i}&=&\vec\sigma_i \times \vec B_i,\label{eq:sieq} \\ \vec B_i&=&(-2g_i \Re(\psi)/\sqrt{N},2g_i\Im(\psi)/\sqrt{N},-E_i),\label{eq:bidef} \end{eqnarray} \endnumparts where
$\vec\sigma_i/2=\vec S_i$ and $\sigma_i^-=\sigma_i^x-i \sigma_i^y$.

We shall describe the model as having excitonic coherence if the phases of the
different $\sigma^-_i$ are correlated. Equivalently, the $xy$ components of the
total spin vector, $\vec\sigma_T=\sum_i \vec\sigma_i$, are of order $N$. This
leads, according to (\ref{eq:ph}), to a state with $\psi \sim\sqrt{N}$.

\subsection{Steady-state solutions}

In previous work\ \cite{eastham03} we identified two classes of
steady-state solutions to (\ref{eq:ph} -- \ref{eq:bidef}) when
$N\to\infty$. The simplest class consists of solutions in which there
is no excitonic coherence and a cavity field of order $N^0$. In this
case as $N\to\infty$ the spin $i$ freely precesses around the $z$ axis
at its natural frequency $E_i$, so the excitons remain incoherent and
the solution is self-consistent. These solutions are generalisations
of the normal state in the equilibrium theory.  When $N$ is finite but
large there is a subset of such solutions which remain incoherent,
including at least those where the cavity field and excitonic
polarisation are exactly zero.

The second class consists of synchronised steady-states which
generalise the condensed solutions of the equilibrium theory. In these
states there is an $O(\sqrt{N})$ component to $\psi$ oscillating at a
single frequency, $\psi \sim \lambda\sqrt{N} e^{-i\mu t}$. In a frame
rotating at this frequency the spin dynamics for large $N$ is just
free precession about a static effective field $\tilde{
  \vec{B}}_i=(-2g_i \Re(\lambda),2g_i \Im(\lambda),-(E_i-\mu))$.
Incorporating this dynamics in (\ref{eq:ph}), we find that the leading
order terms are satisfied if
\begin{equation}
\label{eq:gapsync}
(\omega_c-\mu)\lambda=\frac{\lambda}{N}\sum_i
\frac{g_i^2\sigma^{z\prime}_i(0)}{\sqrt{(E_i-\mu)^2+4g_i^2|\lambda|^2}}.
\end{equation} Here $\sigma_i^{z^\prime}(0)$ denotes the initial component of the $i^{th}$ spin
along its effective field $\tilde{\vec{B}}_i$. If every spin lay
parallel to its effective field $\tilde{\vec{B}}_i$ then
(\ref{eq:gapsync}) would be the gap equation for the polariton
condensate at $T=0$, but we see that there are many more
self-consistent solutions with non-equilibrium distributions of
$\sigma_i^{z^\prime}(0)$.

We note that the self-consistency argument admits the possibility of
steady-states where the spins do not lie along their effective fields,
because the dominant terms in the sum on the right of (\ref{eq:ph})
come from components of $\sigma^-$ which oscillate at frequency $\mu$.
If the spins do not lie along $\tilde{\vec{B}}_i$ then there are
additional components at frequencies $\mu\pm|\tilde{\vec{B}}_i|$, but
since these frequencies differ for each spin they give terms of $O(1)$
in (\ref{eq:ph}). Thus these components are irrelevant when
$N\to\infty$. For $N$ finite but large they lead to noise sources
which drive the $O(1)$ component of $\psi$, and so would not change
the macroscopic behaviour so long as the states are dynamically
stable.

%%  This
%% is not in fact the case for the model with $g_i=g$, where exact
%% solutions (discussed below) show that the steady-states with a
%% constant order parameter have the spins along the effective fields.
%% However, this does not rule out the more general states in the model
%% with a distribution of $g$.

\subsection{Exact solutions}

In recent work\ \cite{yuzbashyan05}, the non-equilibrium gap equation
(\ref{eq:gapsync}) is derived from an exact solution to the model
(\ref{eq:ph} -- \ref{eq:bidef}). This solution applies to the special
case $g_i=g$ in which the model is integrable.  A similar equation is
also given for the steady-states of the BCS model in
\cite{barankov06}. The exact solutions go well beyond the arguments
above, and show that there are many further classes of solutions in
which $|\psi|$ oscillates with a small number of frequencies. The next
simplest solutions correspond to an ansatz proposed by Barankov and
Levitov\ \cite{barankov04}, and have oscillations of the order
parameter described by elliptic functions. An approach for determining
the type of solution which evolves from a given initial condition is
discussed in\ \cite{yuzbashyan06}, and can be applied to
the Dicke model using the Lax vector given in\ \cite{yuzbashyan05}.

\section{Numerical Results}

\label{sec:numerics}

Having seen that the model (\ref{eq:ph} -- \ref{eq:bidef}) has
coherent solutions, we now consider whether the coherence can develop
dynamically when $N$ is finite but large. This question can, for the
integrable model $g_i=g$, be analysed entirely using the exact
solutions. We shall approach it instead through numerical solutions to
the equations of motion, allowing us to go beyond the exactly-solved
model, and return to discuss how the results link to the exact
solution in section\ \ref{sec:exactsols}.

We first consider the model when $g_i=g$, and choose our units of
energy such that $g=2$, and the zero of energy such that
$\omega_c=0$. We take the energies to be drawn from a Gaussian of
standard deviation $\sigma=0.3$ and mean zero, so the photons are
resonant with the centre of the inhomogeneously broadened exciton
line.

The dynamics in general depends on the initial conditions. We consider
initial conditions in which there is no correlation between the spin
and energy on each site. To construct an initial condition with
negligible classical coherence we take the spins to make angles to the
z axis chosen from a uniform distribution over $[\theta_1,\theta_2]$,
and to have angles to the x axis drawn from a uniform distribution
over $[0,2\pi]$. Since we are looking for symmetry-broken solutions we
also include a small initial seed for the cavity field.

\begin{figure}[htpb]
\begin{center}
\includegraphics[width=12cm]{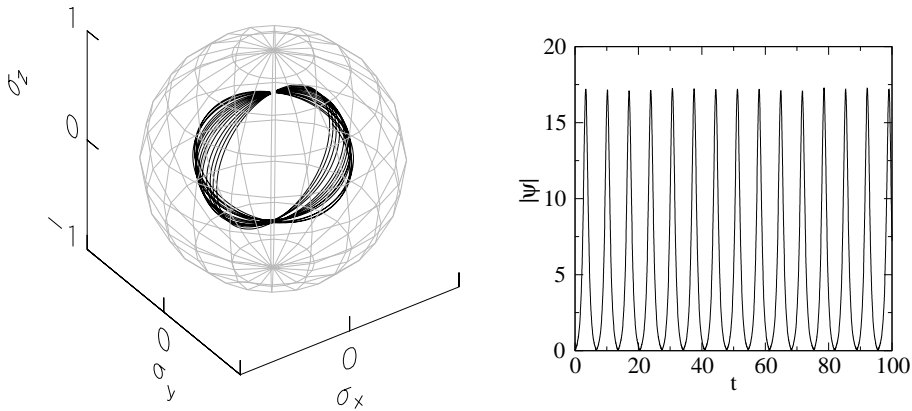}
\caption{Locus of the total spin vector to $t=100$ (left panel) and
  cavity mode amplitude (right panel) for the classical Dicke model
  with $g_i=g=2$ and $N=500$, starting in an unphased inverted initial
  condition. $\omega_c=0$ and the energies $E_i$ are drawn from a
  Gaussian with mean $0$ and standard deviation $0.3$. The initial
  state has a small photon field and spins whose angles to the
  z(x)-axis are drawn uniformly from $[0,\pi]$($[0,2 \pi]$).}
  \label{fig:superradiance}
\end{center}
\end{figure}

For this type of initial conditions our simulations of the dynamics do
not develop coherent photons unless the average $\sigma_i^z(0)$ exceeds a
positive threshold. An example of the behaviour when it does is shown
in figure\ \ref{fig:superradiance}, for $500$ spins with $\theta_1=0$
and $\theta_2=\pi/2$. The left panel shows the locus of the tip of the
total spin vector $\vec\sigma_T / N$ up to $t=100$. The right panel
shows the associated field amplitude $|\psi|$. The behaviour contrasts
with that obtained in a simulation starting from non-inverted initial
conditions, $\sigma^z_T(0)/N<0$, and shown in the left panel of figure\
\ref{fig:damping}. In this case the locus becomes a point as $N\to
\infty$, and the photon field remains of $O(1)$.

As can be seen from figure\ \ref{fig:superradiance}, there are
mechanisms in the Dicke model that can generate coherence from an
incoherent state, even without dissipation. However, for this class of
initial conditions they are only effective in inverted states, whereas
equilibrium condensation does not require inversion. A more
immediately obvious relation is to strong-coupling superradiant
dynamics, demonstrated experimentally in\ \cite{kaluzny83}. This
phenomenon can be treated by considering the classical Dicke model
with $E_i=E$ and $g_i=g$, so that (\ref{eq:ph} -- \ref{eq:bidef})
reduce to equations for $\psi$ and $\vec\sigma_T$. The dynamics of
$\vec\sigma_T$ can then be shown to be that of a rigid
pendulum\ \cite{bonifacio70}. Since the length of $\vec{\sigma}_T$ is
conserved when $E_i=E$ the excitonic system must become coherent if
the pendulum swings towards the equator. The ``gravity'' on this
pendulum acts down in figure\ \ref{fig:superradiance}, so that coherence
increases only from inverted states. This asymmetry in the dynamics is
ultimately due to the rotating-wave approximation made in
(\ref{eq:spinham}), which defines directions of energy flow between
the cavity mode and exciton.

In the absence of a field the polarization of an initially coherent
state undergoes free induction decay, as the different two-level
systems go out of phase due to the inhomogeneous broadening. An
interesting feature of figure\ \ref{fig:superradiance} is that the
coherence does not appear to decay even in the presence of an
inhomogeneous broadening. Such undamped dynamics for the cavity mode
is reminiscent of self-induced transparency\ \cite{mccall69}, in which
soliton-like pulses propagate unattenuated in an inhomogeneously
broadened medium.

It becomes possible to generate spontaneous coherence in the
Hamiltonian dynamics without inversion if one considers initial
conditions in which the spins are correlated with energy. In
particular, Barankov and Levitov recently presented an ansatz
\cite{barankov04} which solves the dynamics starting from a Fermi
sea. In these solutions there is a coherent cavity field whose
amplitude undergoes undamped oscillations like those in figure\
\ref{fig:superradiance}. As we will discuss in section\
\ref{sec:exactsols}, the coherent dynamics we find in simulations
corresponds to the solution given in\ \cite{barankov04}, although our
initial conditions are not considered there.

\begin{figure}[tpb]
\begin{center}
\includegraphics[width=12cm]{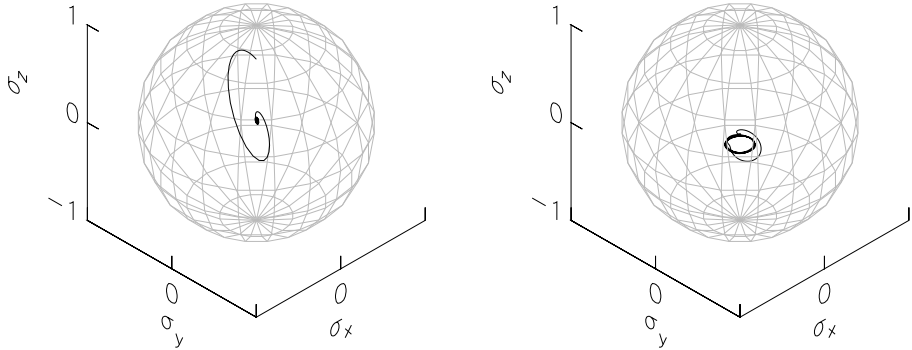}
\caption{Destruction of the gap oscillations by a distribution of
  $g$. Left panel: parameters of figure\ \ref{fig:superradiance} except
  that $N=2000$ and $g$ is drawn from a Gaussian of mean $2$ and
  standard deviation $0.6$. Right panel: same parameters starting from
  a Fermi sea with a Fermi energy $0.05$ below the centre of the
  inhomogeneously broadened exciton line.} \label{fig:distg}
\end{center}
\end{figure}

Although the oscillations are not suppressed by randomness in $E_i$,
it appears that they can be suppressed by randomness in the coupling
constants $g_i$, which exists in a disordered
semiconductor\ \cite{marchetti052}. The left panel of figure\
\ref{fig:distg} shows the dynamics in the Bloch sphere for the same
parameters as figure\ \ref{fig:superradiance}, but with the coupling
constants now drawn from a Gaussian with mean $2$ and standard
deviation $0.6$. The right panel shows the results for the same
parameters but starting from a Fermi sea with the Fermi edge at an
energy $0.05$ below the middle of the band; without the randomness of
$g$ this initial condition results in persistent gap oscillations. In
both cases we see that the oscillations of the macroscopic
electromagnetic field, which according to (\ref{eq:exnum}) correspond
to oscillations in $\sigma^z_T$, disappear at late times. The dynamics
reaches a steady-state condensate with a uniform field, corresponding
to the self-consistency condition (\ref{eq:gapsync}). Starting from an
inverted, uncorrelated state the final solution appears to be the
trivial one in which the locking frequency $\mu=\omega_c$ and there is
no coherent polarization: the two-level systems are in a dephased
state, which has decoupled from a residual cavity field developed
during the early stages of the dynamics. Starting from a Fermi sea,
however, we see that the late-time attractor of the dynamics is a
circle: it has reached non-trivial locked states in which the locking
frequency differs from $\omega_c$ and there is therefore a coherent
excitonic polarization.

\section{Linear stability analysis}

\label{sec:stability}

A simple way to understand why either inversion or a Fermi sea allows
coherence to develop is to consider a linear stability analysis of the
incoherent state. The action for fluctuations around the incoherent
thermal equilibrium state is equation (23) in reference\
\cite{eastham01}, so the real eigenfrequencies $\lambda$ are the
solutions to
\begin{equation}
\omega_c-\lambda-\frac{1}{N}\sum_i \frac{g_i^2 \tanh(\beta \tilde
E_i/2)}{E_i-\lambda}=0, \label{eq:tdnstab}
\end{equation} where $\tilde E_i$ is $E_i$ measured
from the chemical potential for polaritons. More generally one can
linearise the zero-temperature dynamics about an arbitrary normal
state\ \cite{barankov04,littlewood96}, which corresponds to replacing
the thermal equilibrium $\sigma_i^z=-\tanh (\beta \tilde E_i/2)$ in
(\ref{eq:tdnstab}) with the general $\sigma^z_i(0)$ of the non-equilibrium
state :
\begin{equation} \omega_c-\lambda=-\frac{1}{N}\sum_i
\frac{g_i^2 \sigma_i^z(0)}{E_i-\lambda}. \label{eq:nstab}
\end{equation} In passing we note that the analogous
generalization of the thermodynamic action for fluctuations about the
condensate, equation (22) of reference\ \cite{eastham01}, gives the
eigenspectrum for the coherent steady-states with a uniform gap and
every spin parallel or antiparallel to its effective field, but more
generally there are additional terms generated by the components of
the spins transverse to the effective fields.

\begin{figure}[tpb]
\begin{center}
  \includegraphics[width=12cm,bb=20 4 319 140]{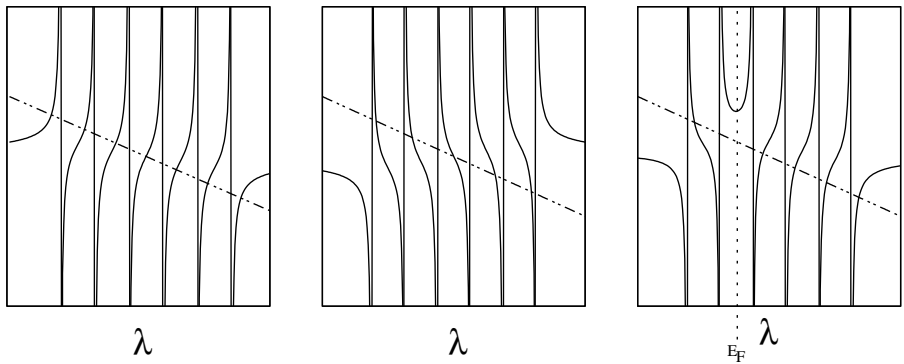}
\caption{Illustration of the left (dot-dashed) and right (solid) sides of the
equation, (\ref{eq:nstab}), determining the eigenfrequencies $\lambda$
of the normal state. Left panel: stable incoherent state with
$\sigma^z_i$ independent of energy and less than zero. Middle panel:
unstable incoherent state with $\sigma^z_i$ independent of energy and
greater than zero. Right panel: unstable Fermi sea, $\sigma^z_i=+(-)1$
for $E_i<(>)E_F$.}
\label{fig:staban}
\end{center}
\end{figure}

A dynamical instability is indicated by the appearance of a complex
root in (\ref{eq:nstab}). In figure\ \ref{fig:staban} we plot the left
and right sides of (\ref{eq:nstab}) for real $\lambda$, a small number
$N=6$ of two-level systems, and three types of initial states. For an
uncorrelated non-inverted state (left panel) there are $N+1$ real roots
which generalise the inhomogeneously broadened polariton spectrum to
the non-equilibrium states. When $g$ becomes large compared with the
bandwidth the two poles outside the band correspond to the usual upper
and lower polariton, with a band of $N-1$ almost excitonic states
between them. The instability of an uncorrelated inverted state is
shown in the middle panel. We see that the two ``polariton'' roots can
become complex, indicating the superradiant instability.

For completeness we illustrate the instability of a Fermi sea in the
right panel. The sign change in $\sigma^z_i(0)$ across the Fermi energy
introduces a turning point in the right-hand side of (\ref{eq:nstab}).
At $T=0$ there can be a stable Fermi sea in the finite system, which
has two ``excitonic'' roots near the Fermi edge. These roots become
complex at the instability, which can be reached by increasing the
system size or changing the detuning. At this instability the pair of
excitons either side of the Fermi edge begin to develop a
polarisation. The generalisation to a Fermi distribution in the
thermodynamic limit is clear: if the distribution has a sharp enough
step, i.e. the temperature is low enough, there can be an instability.

\section{Exact solutions}
\label{sec:exactsols}

In the case $g_i=g$ the model we have considered is exactly solvable,
and its solution is equivalent to that of the BCS model\
\cite{yuzbashyan05}. The linear stability analysis above can be
extended by linking it to the exact solutions through the concept of
the Lax vector. In our notation the Lax vector $\vec{L}(\vec{u})$ for
the generalised Dicke model with $g_i=g$ is\
\cite{yuzbashyan05}\begin{equation}
\tilde{\vec{L}}(\vec{u})=\frac{g^2}{N} \vec{L}(\vec{u})=\left(
\begin{array}{c} 2g \Re{(\psi)}/\sqrt{N} \\ -2g \Im{(\psi)}/\sqrt{N} \\ 2u-\omega_c
\end{array} \right) + \frac{g^2}{N}\sum_i
\frac{\vec{\sigma}_i}{2u-E_i}. \label{eq:lax}\end{equation} As
discussed in\ \cite{yuzbashyan06}, the character of the dynamics is
connected to the number of isolated branch cuts of the conserved
function $\sqrt{\vec{L}^2(\vec{u})}$. For a single cut the long-time
dynamics has a constant order parameter obeying (\ref{eq:gapsync}),
while for two cuts one obtains the oscillating gap solution\
\cite{barankov04}. The particular parameters within each class of
solution, such as order parameters and the locking frequencies, depend
on the initial conditions.

For incoherent initial conditions
$\tilde{\vec{L}}(\vec{u})=\vec{k}L_z(u)+\vec{L}_{xy}(u)$ lies almost
parallel to $\vec{k}$. Comparing (\ref{eq:lax}) with (\ref{eq:nstab})
we see that the zeroes of $L_z(u)$ and the eigenfrequencies of the
linear stability analysis are related as $2u=\lambda$. Thus for the
stable case in the left panel of Fig.\ \ref{fig:staban} the zeros of
$L_z(u)$ lie along the real axis, while in the unstable cases shown
there are two complex conjugate roots off the axis. All these roots
lead to doubly-degenerate roots of $L_z^2(u)$, which are then split by
the perturbation $\vec{L}_{xy}$. Thus the unstable states we consider
have two isolated branch cuts, and thus correspond to the oscillating
gap solution.

\section{Damping}

\label{sec:damping}

\begin{figure}[tpb]
\begin{center}
\includegraphics[width=12cm]{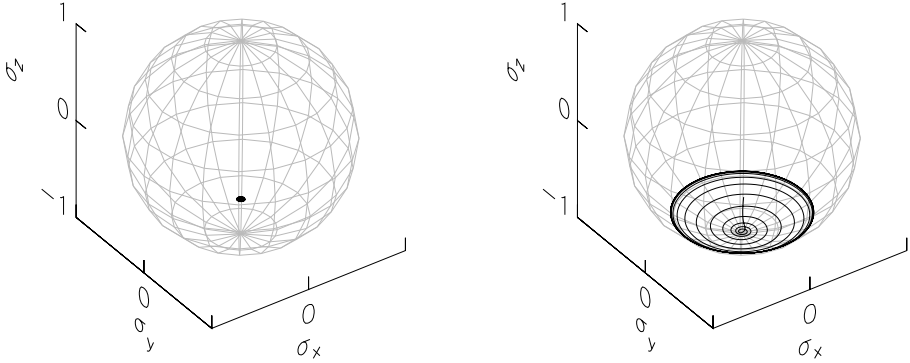}
\caption{Locus of the total spin vector up to $t=100$ starting from an
initial condition where the spins have angles to the x-axis chosen
uniformly from $[0,2\pi]$, and to the z-axis from $[\pi,2\pi]$. Left
panel: undamped model. Right panel: model with phenomenological
damping $\gamma=1$. $N=2000$, $E_i$ from a Gaussian with mean 1 and
standard deviation 0.5, $g_i=g=2$, $\omega_c=1$.} \label{fig:damping}
\end{center}
\end{figure}

We have discussed two dynamical mechanisms for creating coherence in
the undamped Dicke model: an instability of the polariton-like modes
which occurs from inverted initial states, and an instability at the
Fermi edge which occurs from a low-temperature Fermi distribution.
However, since the polariton condensate is the thermal equilibrium
state then at low enough temperatures the mechanisms which establish
thermal equilibrium, i.e. dissipation and scattering, must somehow
generate coherence.

A simple phenomenological way to extend the Dicke model to include
dissipation is to add a term $-\gamma \vec\sigma_i \times (\vec
\sigma_i \times \vec B_i)$ to the right of (\ref{eq:sieq}). By
construction such a term tends to align each spin with its effective
field $\vec B_i$. However the effective field $\vec B_i$ depends on
the choice of rest frame, {\emph{i.e.}} the zero of energy. In the
equilibrium case the spins align with the effective field in a rest
frame rotating at the chemical potential. Thus we see that with the
damping term implicitly introduces a chemical potential, and is
associated with equilibration with an exciton bath.

In the right panel of figure\ \ref{fig:damping} we illustrate the
approach to an equilibrium condensate under the phenomenological
damped dynamics. The initial state here has spins randomly oriented in
the southern hemisphere of the Bloch sphere, and does not develop
coherence without the damping term. The damping term acts to align the
phases of the spins and generate a coherent state. The final locking
frequency of the condensate for these parameters is approximately the
equilibrium chemical potential $\mu_{pol}$ which gives the
steady-state value of $L$; it comes closer still in simulations with
smaller damping terms. $\mu_{pol}$ is not in general identical to the
chemical potential $\mu_{ex}$ implied by the choice of rest frame
because the former is coupled to both excitons and photons and the
latter only to excitons.

\section{Conclusions and Outlook}

\label{sec:conclusions}

In a polariton condensation experiment a microcavity is pumped at high
energies, while with sufficiently strong pumping coherent polaritons
are observed at low energies. A first approximation to this situation
treats the pumping as a high-energy exciton reservoir which randomly
populates the low-energy excitons independently of their
energy. Within this approximation we can suggest several scenarios,
differentiated by the relative timescales for thermalization, exciton
decay and photon decay, which would lead to coherent photons.

If thermalization and exciton decay are negligible we expect a
scenario which might be described as a ``polariton laser''. The
reservoir will build up the inversion until the superradiance
threshold is exceeded, and the dynamical instability serves to create
a coherent photon field. This then opens a decay channel allowing the
excitation $L$ to decay through the photons, which for a large
reservoir is expected to result in a stationary value of $L$ produced
by competition between pumping and damping. 

If thermalization occurs before the excitons decay, and to a low
enough temperature, we could have the analogous scenario but with the
coherence developing before the superradiance threshold. Conceptually
the coherence could develop either through the creation of a Fermi
edge and the resulting dynamical instability or directly through
dissipation, but in practice these are intimately related. An
interesting feature of this scenario is that it suggests that it may
be possible for the dissipation to develop some coherence even when
the dissipation time is longer than the photon lifetime, because the
latter only becomes relevant as the coherence develops.

In conclusion, we have illustrated both dissipative and
non-dissipative mechanisms which generate coherence from incoherent
initial states in Dicke-type models. These include polaritonic and
pairing instabilities, which we have seen generating solutions with
both oscillating and static ordering, and a phenomenological
dissipation mechanism which generates the equilibrium condensate
solution. The non-dissipative mechanisms only lead to coherent
solutions from specific initial conditions. Within the classes of
initial conditions we have considered it appears from our simulations
that either inversion, a Fermi edge, or dissipation are needed to
generate coherence.

\ack

I acknowledge helpful discussions with Peter Littlewood, Jonathan
Keeling, Ben Simons and Francesca Marchetti. This work is supported by
the EPSRC, Sidney Sussex College and the EU network ``Photon-mediated
phenomena in semiconductor nanostructures''.

\section*{References}

%\bibliographystyle{iopart-num}
%\bibliography{classview-reresubmit}

\providecommand{\newblock}{}

\end{document}